\documentclass{article}

\PassOptionsToPackage{numbers, compress}{natbib}

\usepackage{tcolorbox}
\definecolor{cadmiumgreen}{rgb}{0.0, 0.42, 0.24}
\definecolor{oldmauve}{rgb}{0.4, 0.19, 0.28}
\definecolor{royalazure}{rgb}{0.0, 0.22, 0.66}
\definecolor{harvardcrimson}{rgb}{0.79, 0.0, 0.09}
\definecolor{lightmauve}{rgb}{0.86, 0.82, 1.0}
\definecolor{darkbrown}{rgb}{0.4, 0.26, 0.13}%
\usepackage[colorlinks = true,
            linkcolor = royalazure,
            urlcolor  = royalazure,
            citecolor = royalazure,
            anchorcolor = royalazure]{hyperref}

    \usepackage[preprint]{neurips_2024}

\usepackage[utf8]{inputenc} 
\usepackage[T1]{fontenc}    
\usepackage{hyperref}       
\usepackage{url}            
\usepackage{booktabs}       
\usepackage{amsfonts}       
\usepackage{nicefrac}       
\usepackage{microtype}      
\usepackage{xcolor}         
\usepackage{algorithm}
\usepackage{algorithmic}
\usepackage{float}
\usepackage{booktabs}
\usepackage{multirow}
\usepackage{capt-of}
\usepackage{wrapfig,lipsum,booktabs}

\usepackage{enumitem,graphicx}
\usepackage{amsmath,amsfonts,bm}

\def\ie{$i.e.$}

\def\1{\bm{1}}

\def\vtheta{{\bm{\theta}}}

\def\vx{{\bm{x}}}

\DeclareMathOperator*{\Exp}{\mathbb{E}}

\title{Backdoor Mitigation by Distance-Driven Detoxification}

\author{Shaokui Wei\textsuperscript{1} \quad Jiayin Liu\textsuperscript{1} \quad Hongyuan Zha\textsuperscript{1,2}\\
\textsuperscript{1}The Chinese University of Hong Kong, Shenzhen, Guangdong, 518172, P.R. China\\
\textsuperscript{2}Shenzhen Key Laboratory of Crowd Intelligence Empowered Low-Carbon Energy Network
}

\def\Dcl{\mathcal{D}_{cl}}

\def\app{\textbf{Appendix}}

\begin{document}

\maketitle
\begin{abstract}
Backdoor attacks undermine the integrity of machine learning models by allowing attackers to manipulate predictions using poisoned training data. Such attacks lead to targeted misclassification when specific triggers are present, while the model behaves normally under other conditions. This paper considers a post-training backdoor defense task, aiming to detoxify the backdoors in pre-trained models. We begin by analyzing the underlying issues of vanilla fine-tuning and observe that it is often trapped in regions with low loss for both clean and poisoned samples. Motivated by such observations, we propose Distance-Driven Detoxification (D3), an innovative approach that reformulates backdoor defense as a constrained optimization problem. Specifically, D3 promotes the model's departure from the vicinity of its initial weights, effectively reducing the influence of backdoors. Extensive experiments on state-of-the-art (SOTA) backdoor attacks across various model architectures and datasets demonstrate that D3 not only matches but often surpasses the performance of existing SOTA post-training defense techniques.
\end{abstract}    
\section{Introduction}
Over the past decades, Deep Neural Networks (DNNs) have achieved conspicuous progress in various domains and applications, such as face recognition, autonomous driving, and healthcare \citep{he2016deep, liu2020computing, tournier2019mrtrix3, adjabi2020past}. Despite these advancements, DNNs face significant challenges due to their susceptibility to malicious attacks. A notable threat is the rise of backdoor attacks, where adversaries secretly insert backdoors into DNN models during training by subtly altering a subset of the training data. These alterations ensure that the model performs normally on benign inputs while consistently misclassifying inputs with a specific trigger. To protect machine learning systems, especially in high-stakes applications, it is imperative to develop robust defenses against such threats.

To tackle the threats posed by backdoor attacks, researchers have explored a wide range of defense strategies throughout the lifecycle of machine learning systems \cite{wu2023defenses}. This paper specifically focuses on the post-training backdoor defense task, which aims to eliminate backdoors from a given pre-trained models \cite{wubackdoorbench, wu2023defenses, wu2024backdoorbench}. One prominent strategy in this area is fine-tuning the models, \ie, adjusting model weights on an additional dataset, to reduce the impact of backdoor attacks. However, vanilla fine-tuning, which solely employs classification loss on clean samples, has proven insufficient against sophisticated backdoor attacks \cite{wubackdoorbench, wu2023defenses, Zhu_2023_ICCV, sha2022fine}. 

%
One of the primary reasons for the ineffectiveness of vanilla fine-tuning stems from the misalignment between the ideal defense goals and the actual objectives pursued by vanilla fine-tuning. To mitigate backdoor, an ideal objective would aim to minimize the loss of clean data (\textbf{clean loss}) while concurrently increasing the loss of poisoned data (\textbf{backdoor loss}). Due to the inaccessibility of poisoned data to defenders, vanilla fine-tuning focuses solely on minimizing the clean loss, while neglecting the backdoor loss. This oversight renders it ineffective against sophisticated backdoor attacks.
To address such misalignment, a promising research direction is to reconstruct the poisoned samples and fine-tune the model to resist them, thereby explicitly reducing backdoor loss \cite{wang2019neural, zeng2022adversarial, zhu2023neural, wei2024shared}.  While these methods achieve significant performance improvements, they often come with increased computational complexity. Recently, there has been a shift in focus towards refining the fine-tuning process itself. Techniques like sharpness-aware minimization \cite{Zhu_2023_ICCV} have been integrated into the fine-tuning procedure to implicitly reduce backdoor loss, thereby improving its effectiveness for backdoor mitigation. However, more recent findings suggest that even these advanced methods can still be vulnerable to attacks when the optimization process is carefully designed~\cite{huynhforget}.

\begin{figure}[ht]
    \centering
    \includegraphics[width = 0.7\linewidth]{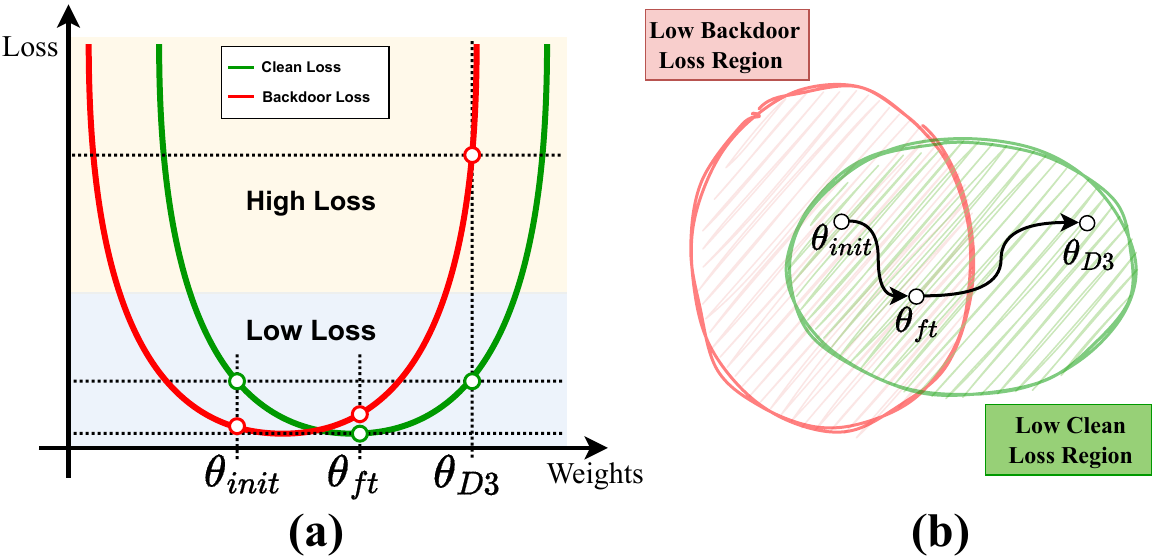}
    \caption{\textbf{(a)}: An illustrative example for curves of clean loss and backdoor loss, highlighting three key points, \ie, the initial weights $\theta_{{init}}$, the weights after vanilla fine-tune $\theta_{{ft}}$, and the weights after applying our method $\theta_{{D3}}$. \textbf{(b)}: A demonstration of loss regions. Vanilla fine-tuning is often trapped in regions where both types of loss are low, thus failing to eliminate backdoors. In contrast, our method finds a more distant solution, thereby escaping this trap and enhancing backdoor defense.}
    \label{fig::landscape}
\end{figure}

In this paper, we delve into the inherent issues of vanilla fine-tuning by investigating the trajectory between the backdoored weights and the fine-tuned weights, revealing that \emph{vanilla fine-tuning often gets trapped in regions with both low clean loss and low backdoor loss}. This phenomenon causes vanilla fine-tuning to converge to bad local solutions, which undermines its defensive efficacy (as illustrated in Figure~\ref{fig::landscape}). To address this challenge, we propose Distance-Driven Detoxification (D3), an innovative strategy for mitigating backdoor attacks. Specifically, we begin by formulating the backdoor defense task as a constrained optimization problem,  with the goal of identifying weights that are maximally distant from the initial backdoored model weights while ensuring that the loss for clean data remains within acceptable bounds. Such goal facilitates the model to escape the region of low backdoor loss, and thereby, effectively mitigating the backdoor effect. Considering the practical challenges, this formulation is subsequently converted into a regularized optimization problem, enabling efficient computation with minimal additional overhead compared to vanilla fine-tuning. Furthermore, we benchmark D3 against eight state-of-the-art (SOTA) post-training defense techniques across seven SOTA backdoor attacks, encompassing various model architectures and datasets. Our experimental results demonstrate that D3 not only achieves performance on par with existing baseline approaches but frequently outperforms them.

Our main contributions are threefold:
\textbf{1) Insight into the failure of vanilla fine-tuning:} We conduct a deeper analysis of the reasons behind the failure of vanilla fine-tuning, revealing that it often becomes trapped in regions with low backdoor loss, which significantly impedes the mitigation of backdoor effects.
\textbf{2) Novel optimization framework:} We introduce a new optimization framework designed to address the limitations of vanilla fine-tuning by maximizing the distance from the backdoored model while maintaining low loss on clean data. This approach provides a straightforward yet effective method to mitigate backdoor attacks.
\textbf{3) Comprehensive evaluation:} We perform extensive experiments to rigorously evaluate the effectiveness of our proposed method, comparing it against eight SOTA defense techniques across seven challenging backdoor attacks, across a diverse set of model architectures and datasets.
\section{Related work}
\paragraph{Backdoor attacks.}
Deep neural networks are vulnerable to backdoor attacks, which pose a significant security threat. These attacks are crafted to ensure that the network performs normally on standard inputs but outputs a pre-defined target when a specific trigger is present. Backdoor attacks can be broadly classified into two categories based on the nature of the trigger: static-pattern backdoor attacks and dynamic-pattern backdoor attacks. The pioneering work in static-pattern backdoors, known as BadNets \citep{gu2019badnets}, utilized fixed triggers such as white squares. To improve the stealthiness of these triggers, the Blended approach \citep{chen2017targeted} was developed, which integrates the trigger seamlessly into the host image. However, the fixed nature of these triggers made them susceptible to detection, leading researchers to shift their focus towards dynamic-pattern backdoor attacks. Recent advancements in dynamic-pattern backdoor attacks include methods like WaNet \citep{nguyen2021wanet}, LF \citep{zeng2021rethinking} and SSBA \citep{li2021invisible}. These techniques generate sample-specific triggers that are more difficult to detect. Additionally, 'clean label' attacks such as LC \citep{shafahi2018poison} and SIG \citep{barni2019new} have been introduced to carry out attacks without disrupting the consistency between the image and its corresponding label. 

\paragraph{Backdoor defenses.}
The primary objective of backdoor defense is to mitigate the susceptibility of DNNs to backdoor attacks. These defenses are generally classified into three categories: pre-training, in-training, and post-training. Pre-training defenses focus on identifying and eliminating poisoned samples before the training process begins. For instance, methods like AC \citep{chen2019detecting}, Confusion Training \citep{qi2023towards}, VDC \citep{zhu2023vdc} leverages various techniques for this purpose. In-training backdoor defenses aim to reduce the impact of backdoors during the training phase. Techniques such as ABL \cite{li2021anti} exploit the faster learning rate of backdoor samples compared to clean samples, and use this to selectively forget poisoned data. DBD \cite{huang2022backdoor} segments the backdoor training process to inhibit the backdoor learning. PDB \cite{wei2024mitigating} and NAB \cite{liu2023beating} eliminate the backdoor effect by injecting another defensive backdoor.

This paper primarily focuses on post-training defenses designed to neutralize backdoors in pre-trained models. One line of research involves identifying and pruning neurons associated with backdoors, including methods like FP \cite{liu2018fine}, ANP \cite{wu2021adversarial}, EP \cite{zheng2022preactivation}, and CLP \cite{zheng2022data}. Another significant approach is to fine-tune the compromised model to mitigate backdoor effects. Notable examples include NC \cite{wang2019neural}, i-BAU \cite{zeng2022adversarial}, NPD \cite{zhu2023neural}, and SAU \cite{wei2024shared}, which use adversarial techniques to reconstruct potential backdoor triggers and then fine-tune the model to resist these reconstructed poisoned samples, effectively cleansing the model. Additionally, NAD \cite{li2021neural} employs a teacher network to guide the fine-tuning process of a backdoored student network, aiding in backdoor mitigation. The work most closely related to our study is FT-SAM \cite{Zhu_2023_ICCV}, which improves the efficacy of fine-tuning for backdoor mitigation through the integration of sharpness-aware minimization. For more detailed information on defenses in adversarial machine learning, readers are directed to the comprehensive survey by \citet{wu2023defenses}.
\section{Method}
In Section \ref{sec::pre}, we establish the notations and introduce the threat model. In Section \ref{sec::rethink}, we revisit vanilla fine-tuning for backdoor defense to identify the underlying issues. Building on these insights, we present a novel optimization formulation for backdoor mitigation in Section \ref{sec::opt}. This formulation can be converted into a regularized optimization problem, facilitating practical and efficient solutions.

\subsection{Problem setting}
\label{sec::pre}

\paragraph{Notations.} 
 In this work, we focus on a classification task aimed at assigning a label \( y \in \mathcal{Y} \) to a sample \( \vx \in \mathcal{X} \). Here, \( \mathcal{Y} = [1, \ldots, K] \) (with \( K \geq 2 \)) represents the set of possible labels, and \( \mathcal{X} \) denotes the sample space. To accomplish this, a model \( f_{\boldsymbol{\vtheta}} \) parametrized by \( \boldsymbol{\vtheta} \) is trained by solving the following problem: \begin{equation} 
 \min_{\boldsymbol{\vtheta}} \mathbb{E}_{(\vx, y) \in \mathcal{D}_{tr}} \left[ \ell(f_{\boldsymbol{\vtheta}}(\vx), y) \right], 
 \end{equation} 
 where \( \mathcal{D}_{tr} \) is the training dataset and \( \ell \) is the loss function. 
 
 In the scenario  of backdoor attacks, the trigger is denoted by \( \Delta \), and a poisoned sample can be created by planting \( \Delta \) to a clean sample \( \vx \), resulting in \( \vx + \Delta \).
\paragraph{Threat model.}
We consider a general backdoor attack scenario where an adversary has the capability to alter a fraction of the training dataset and/or influence the training process to embed a backdoor. The backdoored model behaves normally for benign inputs but misclassifies inputs with the malicious trigger \( \Delta \) to a specific target label \( \hat{y} \). The proportion of altered samples is referred to as the \textbf{poisoning ratio} of the backdoor attack.
\paragraph{Defender's goal.} We consider a post-training backdoor defense scenario. The defender is provided with a model that may have been compromised by a backdoor attack. The primary objective is to mitigate the backdoor so that the model no longer responds to the malicious trigger \( \Delta \) while preserving its performance on the original classification task. We assume the defender has access to a small, clean dataset \( \Dcl \), which is a common assumption in most post-training defense strategies \cite{liu2018fine, li2021neural, wang2019neural,zeng2022adversarial,zhu2023neural,wei2024shared,Zhu_2023_ICCV}. This dataset can be obtained through various means, such as purchasing from trusted data providers, generating using advanced generative models \citep{croitoru2023diffusion, goodfellow2020generative, kingma2013auto}, collecting from online resources, or employing data cleaning techniques \citep{wu2023defenses,chen2019detecting,qi2023towards,zhu2023vdc}. Additionally, the defender does not have prior knowledge of the specific malicious trigger \( \Delta \) or the target label \( \hat{y} \).

\subsection{Revisiting fine-tuning for backdoor mitigation}
\label{sec::rethink}
Given a backdoored model, fine-tuning, which involves adjusting the model weights using an additional dataset, is a promising method for mitigating backdoor effects. Due to its simplicity, vanilla fine-tuning emerges as a natural approach to detoxifying backdoors hidden within the model.
However, previous studies \citep{wubackdoorbench, wu2024backdoorbench, Zhu_2023_ICCV, min2024towards} have observed that vanilla fine-tuning often falls short in effectively removing backdoors. To explore the reason, we first revisit the ideal objective for defending against backdoor attacks. Given a clean dataset $\Dcl$, trigger $\Delta$, a target label $\hat{y}$, and a loss function $\ell$, the ideal fine-tuning objective to purify model $f_{\vtheta}$ is:
\begin{equation}
    \label{eq1}
    \min_{\vtheta} \overbrace{ \underbrace{\Exp_{(\vx,y)\in \Dcl} \left[\ell(f_{\vtheta }(\vx),y)\right]}_{\textbf{Clean loss}}-\underbrace{\Exp_{(\vx,y)\in \Dcl}\left[\ell(f_{\vtheta}(\vx+\Delta),\hat{y})\right]}_{\textbf{Backdoor loss}}}^{\textbf{Ideal loss}}],
\end{equation}
which seeks to preserve the model's performance on clean data while simultaneously unlearning the backdoor. Therefore, an ideally purified model would exhibit both a low clean loss and a high backdoor loss. In contrast, vanilla fine-tuning aims to minimize only the clean loss, which diverges from the ideal objective for backdoor defense. This discrepancy limits the efficacy of vanilla fine-tuning in addressing backdoor threats.

To further understand such discrepancy, we investigate how the backdoor loss evolves along a trajectory connecting the initial model to the fine-tuned model. Starting from a backdoored model with initial weights $\vtheta_{{init}}$, we apply vanilla fine-tuning to obtain fine-tuned weights $\vtheta_{{ft}}$. Then, a trajectory is constructed along the direction $\vtheta_{{ft}} - \vtheta_{{init}}$, where each point on this path is given by $\vtheta_t = \vtheta_{{init}} + t \cdot (\vtheta_{{ft}} - \vtheta_{{init}}).$
We conduct experiments using four  backdoor attacks: BadNets\cite{gu2019badnets}, WaNet \cite{nguyen2021wanet}, Input-aware \cite{nguyen2020input} and SSBA \cite{li2021invisible}, employing the PreAct-ResNet18 architecture. Our primary focus is on vanilla fine-tuning itself; therefore, we fine-tune the backdoored models using all accessible clean samples to eliminate the impact of data insufficiency. We then visualize the changes in loss along the trajectory, with $t$ varying from $0$ to~$2$. 
\begin{figure}[ht]
    \centering
    \includegraphics[width = 0.7\linewidth]{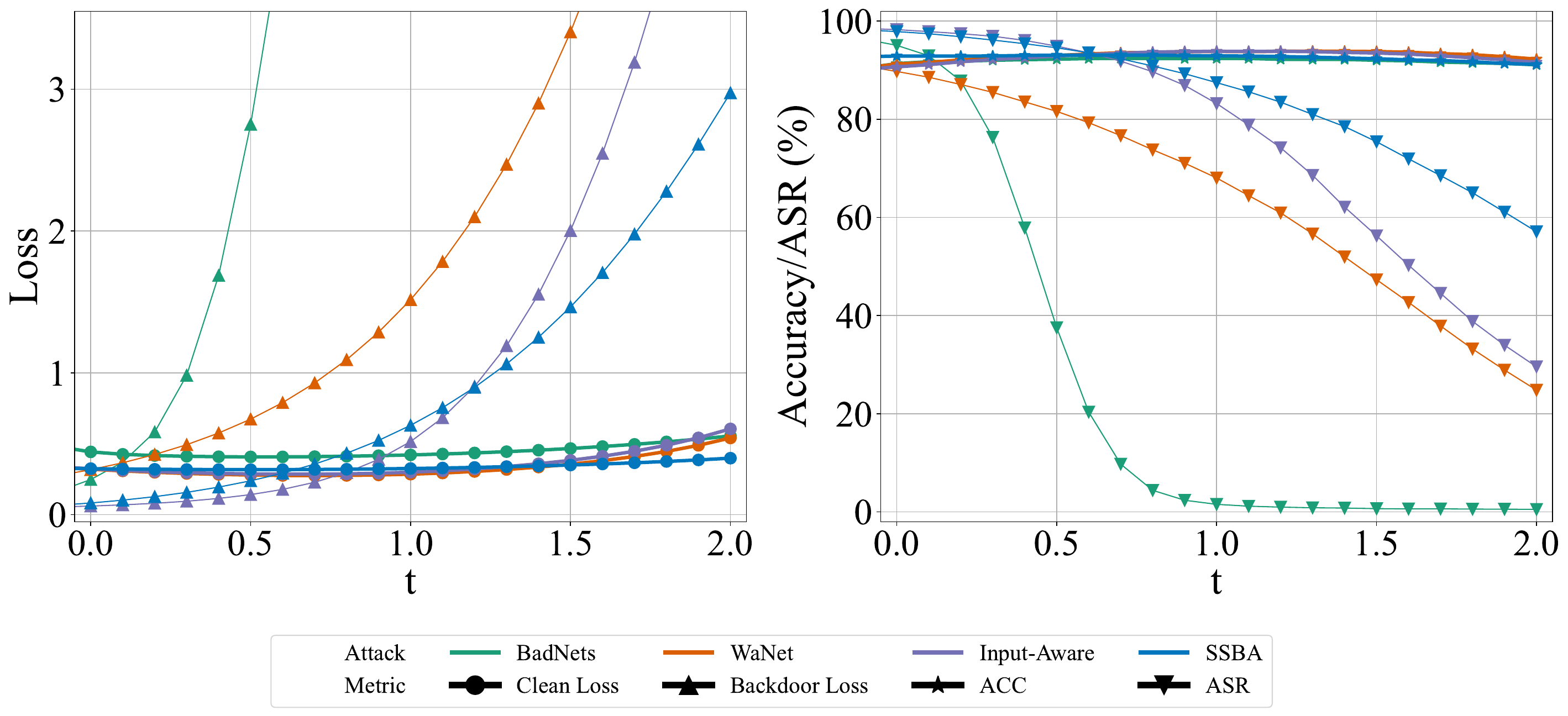}
    \caption{\textbf{Left:} The visualization of curves for clean loss and backdoor loss, along the trajectory for fine-tuning four attacks. \textbf{Right:} The visualization of curves for Accuracy and Attack Success Rate, along the trajectory for fine-tuning four attacks.}
    \label{fig::loss}
\end{figure}

The results are shown in Figure~\ref{fig::loss}, and our key observations are as follows:

\begin{itemize}[leftmargin=*]
    \item \emph{The initial backddoored model are often trapped in a "backdoor region" where vanilla fine-tuning struggles to escape.} When moving from the initial weights ($t=0$) to the fine-tuned weights ($t=1$), we observe a consistent decrease in clean loss, reflecting an improvement in clean accuracy. However, the backdoor loss remains low for most attacks (except for BadNets), indicating that the attack success rate remains high. This implies that vanilla fine-tuning, which focuses solely on clean loss, is insufficient to escape the "backdoor region" characterized by low backdoor loss.
    \vspace{0.01in}
    \item \emph{Shifting the weights away from the initial weights can help mitigate backdoors while preserving clean performance.} By extending the trajectory of the weights slightly beyond the fine-tuned point ($1<t<2$), there is a substantial increase in backdoor loss across various attacks, along with a significant decrease in the Attack Success Rate. Importantly, this adjustment has little negative impact on the clean loss. These findings suggest that increasing the separation between the initial backdoored model and the fine-tuned model is a promising strategy to lower ASR without sacrificing performance on clean data.

\end{itemize}
%
\paragraph{Theoretical justification.} Here, we provide a simple theoretical explanation for the observed behavior. As the initial model is optimized to minimize loss over a poisoned dataset, We assume that \( \vtheta_{init} \) is already a local minimum of the backdoor loss \( \mathcal{L}_{bd} \), \ie, \( \nabla \mathcal{L}_{bd}(\vtheta_{init}) = 0 \). Then, we can  conduct a second-order Taylor expansion of \( \mathcal{L}_{bd}(\vtheta_t) \) about \( \vtheta_{init} \), leading to the following approximation:
\resizebox{\linewidth}{!}{
\begin{minipage}{\linewidth}
\begin{align*}
\mathcal{L}_{bd}(\vtheta_{t}) - \mathcal{L}_{bd}(\vtheta_{init}) &\approx \frac{1}{2} (\vtheta_{t} - \vtheta_{init})^T H(\vtheta_{init}) (\vtheta_{t} - \vtheta_{init}) \\
&= \frac{t^2}{2} (\vtheta_{ft} - \vtheta_{init})^T H(\vtheta_{init}) (\vtheta_{ft} - \vtheta_{init}),
\end{align*}
\vspace{0.05in}
\end{minipage}}
where \( H(\vtheta_{init}) = \nabla^2 \mathcal{L}_{bd}(\vtheta_{init}) \) is the Hessian matrix of \( \mathcal{L}_{bd} \) evaluated at \( \vtheta_{init} \).

Under the above assumption, the second-order conditions for a minimum imply that the loss function curves upwards in every direction. This suggests that the Hessian matrix is positive semi-definite \cite{Liberzon2024}, meaning that for any direction \( u \), the quadratic form \( u^T H(\vtheta_{init}) u \geq 0 \). Therefore, as \( t \) increases, the difference \( \mathcal{L}_{bd}(\vtheta_{t}) - \mathcal{L}_{bd}(\vtheta_{init}) \)  grows approximately quadratically with \( t \). This relationship demonstrates that the backdoor loss \( \mathcal{L}_{bd}(\vtheta_t) \) increases as the weights \( \vtheta_t \) move further away from its initial values, as observed in Figure~\ref{fig::loss}.

\subsection{Distance-driven detoxification}
\label{sec::opt}
\paragraph{Motivation and objective.}
Motivated by the above observations, our goal is to design a strategy that enables fine-tuning to escape the \textit{"backdoor region"} around the initial backdoored weights. A straightforward approach is to identify model weights that are significantly distant from the initial backdoored weights, while still maintaining performance on clean samples. This dual objective can be formalized as a constrained optimization problem:
\begin{equation}
\label{eq::overall} 
\begin{aligned}
    \max_{\vtheta} \quad \quad & d(\vtheta,\vtheta_{init})\\
    \mathrm{s.t.} \quad \quad & \mathbb{E}_{(\vx,y)\in \Dcl} \left[\ell(f_{\vtheta}(\vx),y) \right] \leq \epsilon.
\end{aligned}
\end{equation}
Here, \( d \) denotes the distance metric between the weights $\vtheta$ and the initial backdoored weights \( \vtheta_{{init}} \). The constant $\epsilon\geq0$ serves as a threshold for the model's performance on clean data. A smaller $\epsilon$ imposes a stricter requirement on the model’s accuracy for clean samples, whereas a larger $\epsilon$ allows more flexibility in accuracy to achieve a greater distance from the initial backdoored weights.
\paragraph{Practical challenge.} Despite its promise, Problem~\ref{eq::overall} encounters several practical challenges:

\begin{itemize}[leftmargin=*]
    \item \emph{Challenge 1: Over-fitting.} In post-training backdoor defense,  we typically have access to only a limited reserved dataset. Unlike vanilla fine-tuning, which converges to weights close to the initial values and thereby retains the pre-trained model’s generalization capabilities, our approach aims to find a solution that significantly diverges from the pre-trained model. However, this divergence can weaken the model’s generalization, causing it to over-fit the reserved dataset. To alleviate such an issue, we propose measuring the distance from the initial weights only on a subset of the model's parameters $\vtheta_{s}$, such as those in the final layers. This strategy helps to prevent over-fitting to the reserved dataset, thereby preserving the model’s generalization potential on broader, unseen data.
    \vspace{0.1in}
    \item \emph{Challenge 2: Weights scaling vulnerability.} A significant issue arises from the vulnerability to weight scaling. For example, in a fully connected layer, if the distance from the initial weights is measured using the \( L_1 \) distance, a large distance could be achieved simply by scaling the initial weights up or down. While this may yield a high distance, it has minimal impact on the model’s predictions, as scaling does not affect the \(\arg\max\) operation used for classification. To address this, we impose an additional constraint $\vtheta_{s}\in \mathcal{S}$ to regulate the scale of weights, ensuring that any measured distance corresponds to meaningful changes in the model’s decision-making process rather than trivial transformations that leave little impact on the predictions.  This constraint can be enforced through a projection operator $\mathcal{P}$ that projects the weights onto the space $\mathcal{S}$.
    \vspace{0.1in}    
    \item \emph{Challenge 3: Complexity of the constraint.} Another critical challenge arises from the constraint in Problem~\ref{eq::overall}. Enforcing this constraint requires evaluating the loss of the DNN model over a dataset, which is inherently non-convex and computationally intensive. To make the optimization problem more tractable, we reformulate the problem as a regularized objective, converting the hard constraint into a penalty term. However, it is important to acknowledge that other methods to tackle this challenge might exist and could be explored in future research.
\end{itemize}
\paragraph{Overall process.} Based on the aforementioned analysis, we select a subset of the model's weights, denoted by $\vtheta_{s}$, and define a space $\mathcal{S}$ to constrain these selected weights. In practice, $\vtheta_s$ is usually chosen as weights of one or a few layer and $\mathcal{S}$  is designed to maintain the norm of the selected weights according to specific metrics. We then propose Distance-driven Detoxification (D3), which fine-tunes the model by solving the following optimization problem:
\begin{equation}
\label{eq::d3}
\begin{aligned}
    \min_{\vtheta:\vtheta_{s}\in\mathcal{S}} \quad & -d(\vtheta_s,\vtheta_{init,s}) + \lambda\times \text{max}\left(0,\mathcal{L}_{cl}(\vtheta) - \epsilon\right),
\end{aligned}
\end{equation}
where $\mathcal{L}_{cl}(\vtheta)=\Exp_{(\vx,y)\in \Dcl} \left[\ell(f_{\vtheta}(\vx),y) \right]$ represents the loss on clean samples, and the term $\text{max}\left(0,\mathcal{L}_{cl}(\vtheta) - \epsilon\right)$ acts as a penalty for violating the performance constraint on clean samples: if the clean loss exceeds $\epsilon$, the penalty increases linearly with the excess loss. The hyper-parameter $\lambda\geq0$ controls the trade-off between moving away from the backdoor weights and maintaining performance on clean data. 

To solve this optimization problem, we employ a Projected Gradient Descent (PGD) scheme. Specifically, in each iteration, D3 performs an unconstrained gradient descent step with respect to the minimization objective in Equation \ref{eq::d3}, followed by a projection operation to ensure that $\vtheta_{s}\in \mathcal{S}$. A detailed pseudo-code of the algorithm can be found in \app.

\section{Experiment}
\subsection{Experiment setting}

\paragraph{Backdoor attack.}
To thoroughly assess the effectiveness of our proposed method, we conduct experiments involving seven prominent backdoor attacks: BadNets Attack~\citep{gu2019badnets}, Blended Attack~\citep{chen2017targeted}, WaNet Attack~\citep{nguyen2021wanet}, Low Frequency  (LF) Attack~\cite{zeng2021rethinking}, Input-aware Attack~\citep{nguyen2020input},  Sinusoidal Signal (SIG) Attack~\citep{barni2019new} and Sample-Specific Backdoor Attack (SSBA)~\citep{li2021invisible}. To ensure a fair and consistent evaluation, we use the checkpoints and configurations provided by BackdoorBench \citep{wubackdoorbench, wu2024backdoorbench}, a standardized platform designed for backdoor learning. Specifically, the checkpoints are directly downloaded from the official website~\footnote{\href{http://backdoorbench.com/model_zoo}{http://backdoorbench.com/model\_zoo}} if accessible and generated by the default configuration otherwise. By default, all attacks were executed with a 10\% poisoning rate, targeting the $0^{th}$ class unless otherwise specified. The effectiveness of these backdoor attacks was evaluated on three benchmark datasets: CIFAR-10 \citep{krizhevsky2009learning}, Tiny ImageNet \citep{le2015tiny}, and GTSRB \citep{stallkamp2011german}. We used three different neural network architectures to ensure a comprehensive assessment: PreAct-ResNet18 \citep{he2016identity}, VGG19-BN \citep{simonyan2014very}, and ViT-B-16 \citep{dosovitskiy2020image}. 

\paragraph{Backdoor defense.}
We evaluate our approach against popular and advanced backdoor defense methods, including vanilla fine-tuning, ANP \citep{wu2021adversarial}, Fine-pruning (FP) \citep{liu2018fine}, NC \citep{wang2019neural}, NAD \citep{li2021neural}, i-BAU \citep{zeng2022adversarial} and FT-SAM \cite{Zhu_2023_ICCV} and SAU \citep{wei2024shared}. To ensure a fair comparison, we use the configurations recommended by BackdoorBench \citep{wubackdoorbench, wu2024backdoorbench}. By default, the reserved dataset size is set to 5\% of the training dataset unless otherwise specified. For the proposed method, we set $\vtheta_s$ to be weights of the linear layers, as they is shared across all model architectures. We set $\lambda = 10$ and $\epsilon=0.1$. Moreover, we measure the weight distance by the Frobenius norm of $\vtheta_{s}-\vtheta_{init,s}$ and constrain the Frobenius norm of $\vtheta_{s}$. Due to space limitations, more details about the experiment setups can be found in Appendix 6 and more experiment results, including results for different configuration of the proposed methods,  are presented in Appendix 8.

\paragraph{Metrics.}
To evaluate the effectiveness of each defense method, we employ three primary metrics: Accuracy on Clean Data (\(\textbf{ACC}\)), Attack Success Rate (\(\textbf{ASR}\)), and Defense Effectiveness Rating (\(\textbf{DER}\)). The \(\textbf{ACC}\) measures the model's ability to accurately predict clean samples, whereas the \(\textbf{ASR}\) indicates the percentage of poisoned samples misclassified to the attacker's desired target label. Higher \(\textbf{ACC}\) and lower \(\textbf{ASR}\) values indicate effective backdoor mitigation. The \(\textbf{DER}\), which has been adopted in \citep{Zhu_2023_ICCV, wei2024shared}, is a metric that ranges from 0 to 1 and is designed to measure the tradeoff between the preservation of \(\textbf{ACC}\) and the reduction of \(\textbf{ASR}\). It is calculated using the formula:
\begin{equation}
    \text{DER} = \frac{\max(0, \Delta \text{ASR}) - \max(0, \Delta \text{ACC}) + 1}{2},
\end{equation}
where \(\Delta \text{ASR}\) and \(\Delta \text{ACC}\) denote the reductions in \(\textbf{ASR}\) and \(\textbf{ACC}\) when comparing the defended model to the initial one.

\textbf{Note}: The most effective defense methods exhibit the highest \(\textbf{ACC}\), lowest \(\textbf{ASR}\), and highest \(\textbf{DER}\). In the subsequent experimental results, the top-performing and second-best methods are highlighted in \textbf{boldface} and \underline{underline}, respectively.

\begin{table*}[ht]
\centering
\caption{Results (\%) on CIFAR-10 with PreAct-ResNet18 and poisoning ratio $10.0\%$.}
\label{cifar10_preactresnet18_1}
\scalebox{0.65}{
\begin{tabular}{c|ccc|ccc|ccc|ccc|ccc}
\toprule
Defense $\rightarrow$ & \multicolumn{3}{c|}{No Defense} & \multicolumn{3}{c|}{FT} & \multicolumn{3}{c|}{ANP \cite{wu2021adversarial}} & \multicolumn{3}{c|}{FP \cite{liu2018fine}} & \multicolumn{3}{c}{NC \cite{wang2019neural}} \\ \midrule
Attack $\downarrow$ & \multicolumn{1}{c}{ACC} & \multicolumn{1}{c}{ASR} & \multicolumn{1}{c|}{DER} & \multicolumn{1}{c}{ACC} & \multicolumn{1}{c}{ASR} & \multicolumn{1}{c|}{DER} & \multicolumn{1}{c}{ACC} & \multicolumn{1}{c}{ASR} & \multicolumn{1}{c|}{DER} & \multicolumn{1}{c}{ACC} & \multicolumn{1}{c}{ASR} & \multicolumn{1}{c|}{DER} & \multicolumn{1}{c}{ACC} & \multicolumn{1}{c}{ASR} & \multicolumn{1}{c}{DER} \\ \midrule
BadNets \cite{gu2019badnets} & $91.32$& $95.03$&  N/A & $89.96$& $1.48$& $96.1$& $90.88$& $4.88$& $94.86$& $\underline{91.31}$& $57.13$& $68.95$& $89.05$& $1.27$& $95.75$\\
Blended \cite{chen2017targeted} & $93.47$& $99.92$&  N/A & $92.78$& $96.11$& $51.56$& $92.97$& $84.88$& $57.27$& $\underline{93.17}$& $99.26$& $50.18$& $\textbf{93.47}$& $99.92$& $50.00$\\
WaNet \cite{nguyen2021wanet} & $91.25$& $89.73$&  N/A & $\underline{93.48}$& $17.1$& $86.32$& $91.33$& $2.22$& $93.76$& $91.46$& $1.09$& $\underline{94.32}$& $91.80$& $7.53$& $91.10$\\
LF \cite{zeng2021rethinking} & $93.19$& $99.28$&  N/A & $92.37$& $78.44$& $60.01$& $92.64$& $39.99$& $79.37$& $\textbf{92.90}$& $98.97$& $50.01$& $91.62$& $1.41$& $\underline{98.15}$\\
Input-aware \cite{nguyen2020input} & $90.67$& $98.26$&  N/A & $93.12$& $1.72$& $98.27$& $91.04$& $1.32$& $98.47$& $91.74$& $\textbf{0.04}$& $\textbf{99.11}$& $92.61$& $0.76$& $98.75$\\
SIG \cite{barni2019new} & $84.48$& $98.27$&  N/A & $\underline{90.80}$& $2.37$& $97.95$& $83.36$& $36.42$& $80.36$& $89.10$& $26.20$& $86.03$& $84.48$& $98.27$& $50.00$\\
SSBA \cite{li2021invisible} & $92.88$& $97.86$&  N/A & $92.14$& $74.79$& $61.16$& $\textbf{92.62}$& $60.17$& $68.71$& $\underline{92.54}$& $83.50$& $57.01$& $90.99$& $\underline{0.58}$& $\underline{97.69}$\\
Average & $91.04$& $96.91$&  N/A & $\underline{92.09}$& $38.86$& $78.77$& $90.69$& $32.84$& $81.83$& $91.75$& $52.31$& $72.23$& $90.57$& $29.96$& $83.06$\\

\bottomrule
\toprule

Defense $\rightarrow$ & \multicolumn{3}{c|}{NAD \cite{li2021neural}} & \multicolumn{3}{c|}{i-BAU \cite{zeng2022adversarial}} & \multicolumn{3}{c|}{FT-SAM \cite{Zhu_2023_ICCV}} & \multicolumn{3}{c|}{SAU \cite{wei2024shared}} & \multicolumn{3}{c}{D3 (\textbf{Ours})} \\ \midrule
Attack $\downarrow$ & \multicolumn{1}{c}{ACC} & \multicolumn{1}{c}{ASR} & \multicolumn{1}{c|}{DER} & \multicolumn{1}{c}{ACC} & \multicolumn{1}{c}{ASR} & \multicolumn{1}{c|}{DER} & \multicolumn{1}{c}{ACC} & \multicolumn{1}{c}{ASR} & \multicolumn{1}{c|}{DER} & \multicolumn{1}{c}{ACC} & \multicolumn{1}{c}{ASR} & \multicolumn{1}{c|}{DER} & \multicolumn{1}{c}{ACC} & \multicolumn{1}{c}{ASR} & \multicolumn{1}{c}{DER} \\ \midrule
BadNets \cite{gu2019badnets} & $89.87$& $2.14$& $95.72$& $89.15$& $\underline{1.21}$& $95.83$& $\textbf{91.49}$& $2.28$& $\underline{96.38}$& $88.56$& $1.33$& $95.47$& $90.77$& $\textbf{0.74}$& $\textbf{96.87}$\\
Blended \cite{chen2017targeted} & $92.17$& $97.69$& $50.47$& $87.0$& $50.53$& $71.46$& $92.67$& $11.61$& $93.76$& $90.24$& $\underline{1.57}$& $\underline{97.56}$& $92.29$& $\textbf{0.22}$& $\textbf{99.26}$\\
WaNet \cite{nguyen2021wanet} & $93.17$& $22.98$& $83.38$& $89.49$& $5.21$& $91.38$& $\textbf{93.66}$& $1.31$& $94.21$& $90.32$& $\underline{0.58}$& $94.11$& $93.31$& $\textbf{0.04}$& $\textbf{94.84}$\\
LF \cite{zeng2021rethinking} & $92.37$& $47.83$& $75.31$& $84.36$& $44.96$& $72.75$& $\underline{92.68}$& $6.89$& $95.94$& $90.5$& $\textbf{0.71}$& $97.94$& $92.37$& $\underline{1.31}$& $\textbf{98.57}$\\
Input-aware \cite{nguyen2020input} & $\underline{93.18}$& $1.68$& $98.29$& $89.17$& $27.08$& $84.84$& $\textbf{93.50}$& $1.54$& $98.36$& $91.08$& $0.93$& $98.66$& $92.96$& $\underline{0.06}$& $\underline{99.10}$\\
SIG \cite{barni2019new} & $90.02$& $10.66$& $93.81$& $85.67$& $3.68$& $97.29$& $\textbf{91.13}$& $\underline{0.57}$& $\underline{98.85}$& $88.57$& $1.84$& $98.21$& $89.99$& $\textbf{0.00}$& $\textbf{99.13}$\\
SSBA \cite{li2021invisible} & $91.91$& $77.4$& $59.74$& $87.67$& $3.97$& $94.34$& $92.12$& $3.20$& $96.95$& $90.11$& $\textbf{0.31}$& $97.39$& $91.85$& $0.81$& $\textbf{98.01}$\\
Average & $91.81$& $37.2$& $79.53$& $87.5$& $19.52$& $86.84$& $\textbf{92.46}$& $3.91$& $96.35$& $89.91$& $\underline{1.04}$& $\underline{97.05}$& $91.93$& $\textbf{0.46}$& $\textbf{97.97}$\\
\midrule

\end{tabular}
}
\end{table*}

\subsection{Main results}

\paragraph{Effectiveness of D3.} To validate the effectiveness of D3, we first present the experimental results on the CIFAR-10 dataset. Due to space constraints, additional results for other datasets and model architectures are provided in \app~8. The results clearly demonstrate that D3 effectively mitigates backdoor attacks across various scenarios. Specifically, D3 achieves a notably lower average ASR compared to other methods, indicating its robustness against backdoor attacks. On the CIFAR-10 dataset, D3 attains the second-lowest ASR in six out of seven different types of attacks, with an ASR lower than 1\% for the remaining attack. This consistent performance highlights D3's ability to defend against a wide range of backdoor threats. Similar exceptional performance is observed on the Tiny ImageNet and GTSRB datasets (see \app~8), further emphasizing its effectiveness in reducing the impact of backdoor attacks.

While mitigating backdoor attacks is essential, maintaining high clean accuracy is equally important. Therefore, we also evaluated D3's performance in terms of ACC and DER. It is worth noting that D3 involves pushing the model away from its initial weights to reduce susceptibility to backdoor attacks. This strategy, however, can have a slight negative impact on clean accuracy. Despite this, D3 still manages to achieve a slightly lower clean accuracy compared to the best-performing baseline. Notably, D3 achieves the best DER in most cases, which measures the balance between ACC and ASR. This indicates that D3 not only effectively reduces the risk of backdoor attacks but also maintains a balanced performance in terms of overall model accuracy.

\paragraph{Influence of poisoning ratio.}
To thoroughly investigate the impact of the poisoning ratio on the effectiveness of the D3, we conduct a series of experiments where the poisoning ratio varies from 1\% to 50\%. This range is chosen to cover a spectrum from minimal to significant poisoning scenarios, allowing us to understand how D3 performs under different levels of adversarial attack intensity. Our findings, as shown in Table \ref{ratio}, reveal that D3 maintains a high level of robustness and effectiveness even when confronted with substantial data poisoning (up to poisoning ratio 50\%). 
\begin{table}[H]
\centering
\caption{Results  (\%) under different poisoning ratio on CIFAR-10 and PreAct-ResNet18.}
\label{ratio}
\scalebox{0.8}{
\begin{tabular}{c|cc|cc|cc}
\toprule
Poisoning ratio $\rightarrow$   & \multicolumn{2}{c|}{1.00\%} & \multicolumn{2}{c|}{10\%} & \multicolumn{2}{c}{20\%} \\ \midrule
Attack $\downarrow$ & ACC          & ASR         & ACC         & ASR        & ACC         & ASR        \\ \midrule
BadNets \cite{gu2019badnets} & 92.18        & 0.68        & 90.77       & 0.74       & 89.94       & 0.77       \\
Blended \cite{chen2017targeted} & 92.85        & 0.24        & 92.99       & 0.22       & 91.95       & 0.02       \\
LF \cite{zeng2021rethinking}      & 92.32        & 1.28        & 92.37       & 1.31       & 91.38       & 0.13       \\ \bottomrule \toprule
Poisoning ratio $\rightarrow$    & \multicolumn{2}{c|}{30\%}   & \multicolumn{2}{c|}{40\%} & \multicolumn{2}{c}{50\%} \\ \midrule
Attack $\downarrow$  & ACC          & ASR         & ACC         & ASR        & ACC         & ASR        \\ \midrule
BadNets \cite{gu2019badnets} & 88.97        & 0.99        & 88.45       & 1.02       & 86.90       & 1.51       \\
Blended \cite{chen2017targeted} & 90.80        & 0.09        & 89.84       & 0.03       & 89.01       & 0.03       \\
LF \cite{zeng2021rethinking}      & 90.67        & 0.08        & 90.02       & 0.61       & 88.91       & 2.10      \\ \bottomrule
\end{tabular}}
\vspace{-0.1in}
\end{table}

\paragraph{Influence of reserved dataset.}
We examine the influence of the reserved dataset from two angles: the size of the reserved dataset and its source. Regarding the size of the reserved dataset, we assess the performance of the proposed method under varying reserved dataset size, ranging from \(1\%\) to \(10\%\) of the training dataset. The results are presented in Table \ref{clean}, from which we can find that D3 remains effective in defending against backdoor attacks even with a small reserved dataset size. However, the ACC declines as the size of the reserved dataset diminishes, a challenge we previously discussed in Section \ref{sec::opt}. 
\begin{table}[ht]
\centering
\caption{Results (\%) on D3 with different reserved dataset size.}
\label{clean}
\scalebox{0.8}{
\begin{tabular}{c|cc|cc|cc}
\toprule
Attack $\rightarrow$  & \multicolumn{2}{c|}{BadNets \cite{gu2019badnets}}                        & \multicolumn{2}{c|}{Blended \cite{chen2017targeted}}                       & \multicolumn{2}{c}{WaNet \cite{nguyen2021wanet}}                          \\ \midrule
Reserved size $\downarrow$ & {ACC} & {ASR} & {ACC} & {ASR} & {ACC} & {ASR}        \\ \midrule
1.0\%    & 88.57                   & 2.31                    & 90.64                   & 2.86                    & 91.96                   & 1.42                       \\
5.0\%    & 90.77                   & 0.74                    & 92.29                   & 0.22                    & 93.31                   & 0.04                    \\
10.0\%   & 90.97                   & 0.44                    & 92.61                   & 0.01                    & 93.53                   & 0.11                   \\ \bottomrule
\end{tabular}}
\end{table}

To facilitate the practical application of D3, one potential approach is to leverage advanced generative models for sample collection. To test our method in such a scenario, we conducted experiments using 2,500 samples drawn from the CIFAR-5m dataset \cite{nakkiran2020deep}, which consists of synthetic CIFAR-10-like images created using the Denoising Diffusion Probabilistic Model \cite{ho2020denoising}. The outcomes of these experiments are summarized in Table \ref{fake}, indicating that D3 continues to perform effectively when utilizing a generated dataset.
\begin{table}[ht]
\centering
\caption{Results(\%) using generated dataset CIFAR-5m.}
\label{fake}
\scalebox{0.8}{
\begin{tabular}{c|cc|cc|cc}
\toprule
Attack $\rightarrow$    & \multicolumn{2}{c|}{BadNets \cite{gu2019badnets}} & \multicolumn{2}{c|}{Blended \cite{chen2017targeted}} & \multicolumn{2}{c}{WaNet \cite{nguyen2021wanet}} \\ \midrule
Data  $\downarrow$      & ACC          & ASR         & ACC           & ASR         & ACC            & ASR           \\ \midrule
CIFAR-5m & 90.42        & 1.11        & 92.16         & 0.20        & 92.85          & 0.04          \\ \bottomrule
\end{tabular}}
\end{table}

\begin{figure}[ht]
    \centering
    \includegraphics[width = 0.5\linewidth]{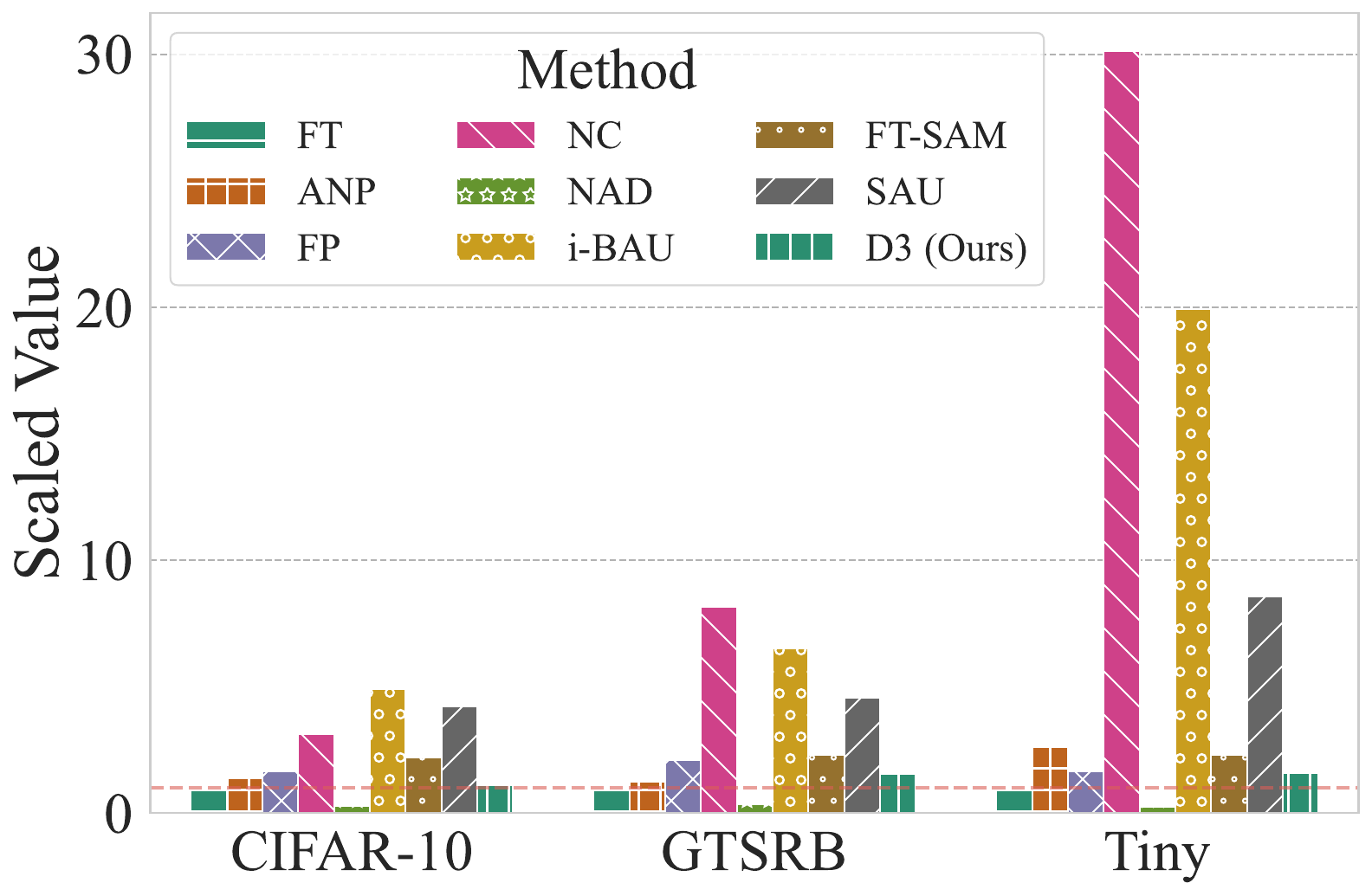}
    \caption{Comparative analysis of different methods across various Datasets. The values are normalized with respect to the training time of vanilla full-tuning.}
    \label{fig::time}
\end{figure}
\paragraph{Training cost comparison.}
Compared to vanilla fine-tuning, D3 incorporates the computation of the weight distance, which incurs only a minimal increase in computational cost. To underscore the efficiency of D3, we have benchmarked its execution time against other baselines. The experiments are performed with CIFAR-10, utilizing PreAct-ResNet18. All tests were executed on a server with an RTX 3090 GPU and an AMD EPYC 7543 32-Core Processor, ensuring a consistent number of training epochs across all methods for a fair evaluation. As illustrated in Figure~\ref{fig::time}, D3 exhibits faster execution times compared to most other defense strategies. 

\subsection{Understanding D3}

\begin{figure}[ht]
    \centering
    \includegraphics[width = 0.6\linewidth]{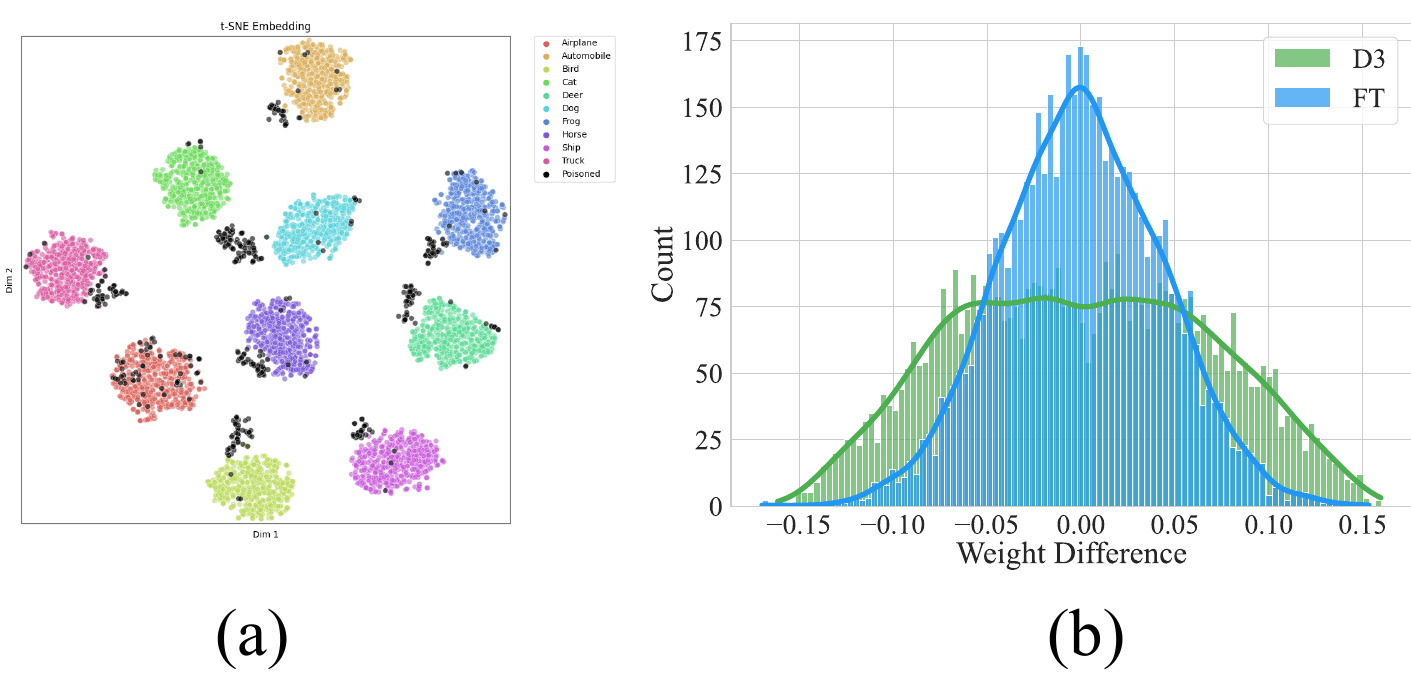}
    \caption{\textbf{(a)} The T-SNE visualization of the model after applying D3 against BadNets attack. \textbf{(b)} Visualization of the weight differences for each parameter in the selected layer. The blue part represents the difference between the vanilla fine-tuned model and the backdoor model, while the green part shows the difference between the model after applying D3 and the backdoor model.}
    \label{fig::loss2}
\end{figure}

In this section, we delve deeper into the mechanism of D3. Firstly, we present the T-SNE visualization (Figure~\ref{fig::loss2}~\textbf{(a)}) of the model after applying D3, which demonstrates that our proposed method can effectively alleviate the backdoor effect, thereby allowing poisoned samples to revert to their original clusters. Additionally, we examine the model from the perspective of weight distances, offering a histogram visualization (Figure~\ref{fig::loss2}~\textbf{(b)}) that highlights the differences in weights. This visualization reveals that the weight difference between the model after D3 and the initial model is notably greater than that between the vanilla fine-tuned model and the initial model. This significant increase in weight difference shows that D3 can effectively find a more distant solution, thereby mitigating the backdoor effect.

\subsection{Resistance to adaptive attack}
In our earlier experiments, we assume that attackers are unaware of the defense method. However, when attackers are aware of the deployment of D3, they could devise adaptive strategies to bypass D3. Given that D3 reduces the impact of backdoors by shifting the model's weights away from their original values, an intuitive counter-strategy for attackers would be to reinforce the backdoor by refining the backdoored weights towards a flat minimum, thereby making it more challenging for D3 to eliminating the backdoor \cite{huynhforget}. To implement this adaptive attack, we adopt Sharpness-aware Minimization (SAM) techniques \cite{sun2023adasam,kwon2021asam}, searching for a flat minimum for the backdoored model as suggested in the work of \citet{huynhforget}.

To assess the resilience of our method against adaptive attacks, we performed evaluations using backdoor attacks with a range of perturbation budgets for SAM, varying from $1.0$ to 
$3.0$. These budgets represent different levels of flatness in the loss landscape that attackers could potentially exploit. As shown in Table~\ref{sam}, both FT and FT-SAM struggle to escape the backdoor region as the flatness increases, ultimately converging on weights with high ASR. Conversely, D3 effectively counters backdoor attacks at all tested levels of flatness by actively steering the model away from backdoored weights. These findings underscore D3's robustness in mitigating backdoor threats.

\begin{table}[ht]
\centering
\caption{Defense results (\%) against adaptive attacks on CIFAR-10 and PreAct-ResNet18.}
\label{sam}
\scalebox{0.7}{
\begin{tabular}{c|c|cc|cc|cc}
\toprule
\multicolumn{2}{c|}{Budget $\rightarrow$ }            & \multicolumn{2}{c|}{1.0}                             & \multicolumn{2}{c|}{2.0}                             & \multicolumn{2}{c}{3.0}                             \\ \midrule
Attack $\downarrow$                   & Defense $\downarrow$   & {ACC} & {ASR} & {ACC} & {ASR} & {ACC} & {ASR} \\ \midrule
\multirow{4}{*}{BadNets \cite{gu2019badnets}} & No Defense & 90.42                   & 96.43                   & 90.71                   & 96.28                   & 91.56                   & 95.60                   \\
& FT         & 92.02                   & 26.30                   & 92.64                   & 33.01                   & 92.55                   & 71.24                   \\
& FT-SAM \cite{Zhu_2023_ICCV}     & 91.28                   & 17.74                   & 92.01                   & 36.34                   & 92.19                   & 54.79                   \\
& D3 (\textbf{Ours})         & 90.90                   & 0.76                    & 91.81                   & 0.72                    & 91.53                   & 1.24                    \\ \midrule
\multirow{4}{*}{Blended \cite{chen2017targeted}} & No Defense & 89.77                   & 99.73                   & 91.30                   & 99.37                   & 91.76                   & 99.78                   \\
& FT         & 92.02                   & 71.80                   & 92.68                   & 83.42                   & 92.46                   & 82.17                   \\
& FT-SAM \cite{Zhu_2023_ICCV}     & 92.17                   & 72.71                   & 92.14                   & 82.80                   & 92.39                   & 91.93                   \\
& D3 (\textbf{Ours})         & 91.13                   & 0.14                    & 91.56                   & 1.22                    & 91.49                   & 2.74                    \\ \midrule
\multirow{4}{*}{WaNet \cite{nguyen2021wanet}}   & No Defense & 89.74                   & 91.52                   & 90.49                   & 90.70                   & 90.36                   & 96.13                   \\
& FT         & 91.71                   & 15.88                   & 92.00                   & 18.47                   & 91.90                   & 21.38                   \\
& FT-SAM \cite{Zhu_2023_ICCV}     & 91.67                   & 5.57                    & 92.19                   & 10.97                   & 92.10                   & 18.87                   \\
& D3 (\textbf{Ours})         & 90.98                   & 0.84                    & 91.00                   & 0.99                    & 90.89                   & 1.48                   \\ \bottomrule
\end{tabular}}
\end{table}

\subsection{Ablation Study}
\label{sec:abl}

\paragraph{Investigation of threshold $\epsilon$.} 
Recall that $\epsilon$ represents the constraint on the loss for clean samples. A lower value of $\epsilon$ indicates a stricter requirement for clean sample accuracy. Specifically, if the cross-entropy loss is used, a loss lower than $\epsilon$ implies a prediction confidence  higher than $\exp(-\epsilon)$. To understand the effect of $\epsilon$, we evaluate D3 with $\epsilon$ values ranging from 0 to 0.5. The results are summarized in Table \ref{eps}. From the table, we observe that a lower $\epsilon$ leads to a higher ACC but also a higher ASR. Notably, even when $\epsilon$ is set to 0, D3 still achieves acceptable defense performance.

\begin{table}[H]
\centering
\caption{Results (\%) on D3 with different thresholds $\epsilon$.}
\label{eps}
\scalebox{0.75}{
\begin{tabular}{c|cc|cc|cc|cc}
\toprule
Attack $\rightarrow$  & \multicolumn{2}{c|}{BadNets \cite{gu2019badnets}} & \multicolumn{2}{c|}{Blended \cite{chen2017targeted}} & \multicolumn{2}{c|}{WaNet \cite{nguyen2021wanet}} & \multicolumn{2}{c}{Average} \\ \midrule
$\epsilon$ $\downarrow$ & ACC          & ASR         & ACC           & ASR         & ACC          & ASR        & ACC           & ASR         \\ \midrule
0       & 91.06        & 1.13        & 92.42         & 1.29        & 93.40        & 0.10       & 92.29         & 0.84        \\
0.1     & 90.77        & 0.74        & 92.29         & 0.22        & 93.31        & 0.04       & 92.12         & 0.33        \\
0.2     & 90.52        & 0.77        & 92.26         & 0.11        & 93.19        & 0.02       & 91.99         & 0.30        \\
0.3     & 90.14        & 0.63        & 92.37         & 0.07        & 93.17        & 0.01       & 91.89         & 0.24        \\
0.4     & 90.32        & 0.53        & 92.34         & 0.07        & 93.13        & 0.00       & 91.93         & 0.20        \\
0.5     & 90.31        & 0.48        & 92.25         & 0.06        & 93.07        & 0.01       & 91.88         & 0.18        \\ \bottomrule
\end{tabular}}
\end{table}

\paragraph{Investigation of multiplier $\lambda$.}
As $\lambda$ is the multiplier for the penalty term, higher  $\lambda$ indicates a stronger emphasis on maintaining high clean sample accuracy. To explore the effect of $\lambda$, we evaluated D3 with $\lambda$ values ranging from 1 to 40. The results are presented in Table \ref{lmd}, which show that a higher $\lambda$ results in a higher ACC but also a higher ASR.

\begin{table}[H]
\centering
\caption{Results (\%) on D3 with different multipliers $\lambda$.}
\label{lmd}
\scalebox{0.75}{
\begin{tabular}{c|cc|cc|cc|cc}
\toprule
Attack $\rightarrow$  & \multicolumn{2}{c|}{BadNets \cite{gu2019badnets}} & \multicolumn{2}{c|}{Blended \cite{chen2017targeted}} & \multicolumn{2}{c|}{WaNet \cite{nguyen2021wanet}} & \multicolumn{2}{c}{Average} \\ \midrule
$\lambda$ $\downarrow$    & ACC          & ASR         & ACC          & ASR          & ACC          & ASR        & ACC           & ASR         \\ \midrule
1 & 88.66        & 0.19        & 90.18        & 0.13         & 91.21        & 0.16       & 90.02         & 0.16        \\
10& 90.77        & 0.74        & 92.29        & 0.22         & 93.31        & 0.04       & 92.12         & 0.34        \\
20& 91.14        & 1.41        & 92.70        & 1.17         & 93.35        & 0.09       & 92.40         & 0.89        \\
30& 91.36        & 2.03        & 92.57        & 4.28         & 93.32        & 0.49       & 92.42         & 2.27        \\
40& 91.34        & 2.01        & 92.78        & 23.36        & 93.35        & 2.84       & 92.49         & 9.40         \\ \bottomrule
\end{tabular}}
\end{table}

\textbf{In summary}, both $\epsilon$ and $\lambda$ have significant impacts on the performance of D3. Lowering $\epsilon$ and increasing $\lambda$ both enhance the ACC, but they also increase the ASR, showing a trade-off between defense effectiveness and model accuracy.

\textbf{Note:} Due to space constraints, further ablation studies on other components of D3 are deferred to \app~7.

\section{Conclusion}

In this work, we delve into the limitations of vanilla fine-tuning for mitigating backdoor attacks in a post-training context. Our analysis shows that vanilla fine-tuning frequently becomes trapped in regions with low loss for both clean and poisoned samples, which hinders effective backdoor removal. To address this challenge, we propose Distance-Driven Detoxification, a novel method that formulates backdoor defense as a constrained optimization problem. D3 encourages the model to move away from its initial weight space, thereby diminishing the impact of backdoors. Comprehensive experimental results across SOTA various ackdoor attacks, multiple model architectures, and diverse datasets confirm that D3 not only matches but often outperforms existing SOTA post-training defense methods, rendering it a promising method for backdoor mitigation.

\paragraph{Limitations and future work.}  One important direction for future work, and a current challenge,  is addressing the trade-off between Accuracy and Attack Success Rate. One promising solution to alleviate such issue involves identifying backdoor-related weights, which can facilitate more accurate distance measurements and help mitigate the trade-off between ACC and ASR. Additionally, extending the D3 framework to handle more complex and diverse attack scenarios is another valuable direction. 

\paragraph{Structure of \app.} The details of the experiments are provided in \app~6. More discussion and analysis, such as connection and comparison to more methods, more ablation study are provided in \app~7. Additional experiments on different settings such as datasets, model structures, poisoning ratios are given in \app~8. 

\newpage
\bibliographystyle{plainnat}
\bibliography{references}

\end{document}